\documentclass[prl, final, twocolumn,amssymb]{revtex4}

\usepackage{amsmath} 
\usepackage{amssymb} 
\usepackage{longtable}
\usepackage{epsfig} \usepackage{bm}

\begin{document}
\title{Fragile-to-Strong Crossover in Supercooled Liquids Remains Elusive}
\author{Yael S. Elmatad}
\email{yael.elmatad@berkeley.edu}
\affiliation{Department of Chemistry, University of California, Berkeley, CA 94720, USA}
\date{\today}

\maketitle

Transport properties of glass forming liquids change markedly around an onset temperature $T_\mathrm{o}$.  For temperatures $T$ above $T_\mathrm{o}$, these properties depend little with $T$, while for $T< T_\mathrm{o}$ these properties are \emph{super}-Arrhenius -- varying \emph{faster} than exponentially in $1/T$.  Upon lowering temperature significantly further, theory \cite{DCJPPNAS} predicts that reversible transport in a supercooled liquid, if it can be observed, will ultimately cross over from super-Arrhenius to Arrhenius temperature variation.  But this so-called ``fragile-to-strong'' (FS) crossover at a temperature $T_\mathrm{x} < T_\mathrm{o}$ has proved difficult to observe because most bulk fluids fall out of equilibrium at a glass transition temperature $T_\mathrm{g}$ that is higher than $T_\mathrm{x}$.  Yet Mallamace et al  \cite{Mallamace2010} report the observation of $T_\mathrm{x}$ for a large number of supercooled liquids.  In truth, they observe the onset to supercooled behavior, and the reported values of $T_\mathrm{x}$ are poor lower-bound estimates to onset temperatures $T_\mathrm{o}$.  This fact is consistent with transport decoupling appearing only below temperatures identified with the crossover in Ref.~\cite{Mallamace2010}.

\begin{figure}
\begin{center}
\resizebox{3.5in}{!}{\includegraphics{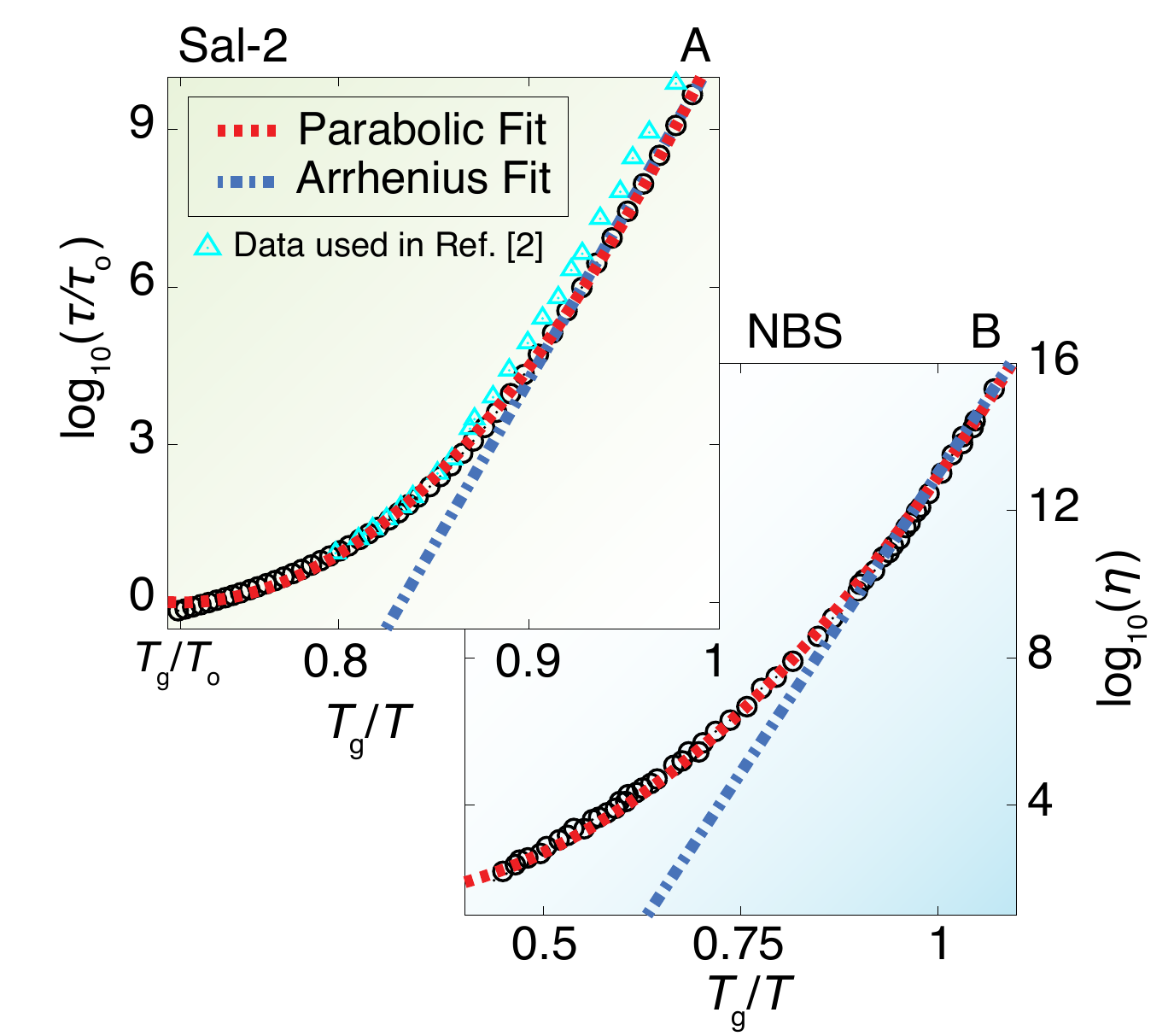}}
\caption{\label{fig:crossover}  Transport properties as a function of $T_{\mathrm{g}}/T$ for two typical supercooled liquids.  Black circles in Panels \emph{A} and \emph{B} represent experimental data considered in \cite{Corr_1}.  Labeling here is consistent with \cite{Corr_1} - that is to say that Sal-2 and NBS  refer to the same experimental measurements and fit parameters as in Table 1 of \cite{Corr_1}.  Red dashed line is the fit parabolic form for $T<T_{\mathrm{o}}$, as in \cite{Corr_1}.  Blue dashed line represents Arrhenius fit for lowest $T$ points \cite{Mallamace2010}. In Panel \emph{A}, relaxation time, $\tau$, of Salol where $T_{g}$ = 221 K is the glass transition temperature where $\log (\tau_\mathrm{g}/\text{s}) = 2$.  In Panel \emph{B}, viscosity, $\eta$, of NBS where $T_{g}$ = 708 K is the glass transition temperature where $\log (\eta_\mathrm{g}/\text{Poise}) = 13$.  It is generally assumed that $\tau \propto \eta$, and, with this assumption, Panel \emph{A} includes data used in Ref. \cite{Mallamace2010} (triangles).
}
\end{center}
\end{figure}

To illustrate this understanding about Ref. \cite{Mallamace2010}, I graph data in Fig.~\ref{fig:crossover} for two typical supercooled liquids.   The data is compared with the parabolic form for transport property $\tau$ (denoting either relaxation time or viscosity),
\begin{equation}
\log_{10} \tau = \log_{10} \tau_{\mathrm{o}} + J^{2}\left(\frac{1}{T}-\frac{1}{T_\mathrm{o}} \right)^{2}, \quad T<T_{\mathrm{o}}\ \,,
\label{parab}
\end{equation}
and with the straight-line fit that would be associated with the Arrhenius temperature dependence.  I consider data for liquid salol in my Fig. \ref{fig:crossover}\emph{A}.  This is the same liquid and the same temperature range considered in Fig. 1\emph{A} of Ref.~\cite{Mallamace2010}.  The two figures are strikingly different, due in part to Ref.~\cite{Mallamace2010} showing an outlying data set \cite{Laughlin} that is discredited by subsequent studies on the same liquid \cite{Corr_1}. 
Unlike Fig. 1\emph{A} of Ref.~\cite{Mallamace2010}, my graph shows excellent agreement between reproducible experimental data and the parabolic form.

My Fig.~\ref{fig:crossover}\emph{B} considers a second liquid to illustrate that the behavior for salol is consistent with that of other systems.  Indeed, the parabolic form, with its three material properties $\tau_\mathrm{o}$, $J$ and $T_\mathrm{o}$, has been used to collapse data for more than 50 supercooled liquids \cite{Corr_1} over the entire supercooled temperature range, $T_\mathrm{o} > T > T_\mathrm{g}$, and this form appears to be universal for all fragile glass formers \cite{Corr_2}. I have chosen to show two specific examples in this Letter to contrast with the obscuring clutter of data analyzed with six-parameter fits in Figs. 1\emph{C} and 2 of Ref.~\cite{Mallamace2010}.  

The two graphs presented here and those presented in Refs.~\cite{Corr_1} and~\cite{Corr_2} indicate that all reliable reversible transport data for bulk supercooled liquids appear to be smooth, with no compelling feature suggesting a change from parabolic to linear behavior.  Rather, it seems that the FS crossover reported in \cite{Mallamace2010} results from confusing $T_\mathrm{o}$ with $T_\mathrm{x}$, and how, over a limited range, a parabola looks like a straight line. The search for the FS crossover in bulk materials therefore remains elusive.    
\section{Acknowledgements}
This work was supported by an NSF GRFP Fellowship for Y.S.E.

\end{document}